\documentclass{amsart}
\usepackage{graphics}

\usepackage{amsmath}

\usepackage{dcolumn}

\DeclareMathOperator{\Spin}{Spin}
\DeclareMathOperator{\SU}{SU}

\newcommand{\cxymatrix}[1]{\vcenter{\xymatrix{#1}}}

\input xy \xyoption{all} \xyoption{poly} 
\labelmargin-{1.2pt}  

\input{epsf}             
 
\newcommand{\tr}{{\rm tr}}

\newcommand{\maps}{\colon} 
\newcommand{\tensor}{\otimes}

\newcommand{\FourX}[4]  
{  
\cxymatrix{\ar@{-}[dr]^{#1} & & \ar@{-}[dl]_{#2} \\  
 & *{\bullet} & \\  
\ar@{-}[ur]_{#3} & & \ar@{-}[ul]^{#4}\\}  
}  
 
\newcommand{\DoubleY}[5] 
{  
\cxymatrix{\ar@{-}[dr]^{#1} & & \ar@{-}[dl]_{#2} \\  
 & *{\bullet} \ar@{-}[d]^{#5} \\
 & *{\bullet} \\
\ar@{-}[ur]_{#3} & & \ar@{-}[ul]^{#4}}  
}  

\newcommand{\DoubleYhor}[5] 
{  
\xymatrix{\ar@{-}[dr]^{#1} & & & \ar@{-}[dl]_{#2} \\  
 & *{\bullet} \ar@{-}[r]^{#5}& *{\bullet} &\\  
\ar@{-}[ur]_{#3} & & & \ar@{-}[ul]^{#4}\\}  
}  
 
\newcommand{\monogon}[1]  
{  
\rule[-1ex]{0mm}{4ex} 
\xymatrix{ *{\bullet}  
\ar@  
{-}  
@(ul,dl)  
[]  
_{#1}  
\\}  
}  
 
\newcommand{\bigon}[2]  
{  
\xymatrix{ *{\bullet} 
\ar@{-} 
@/^1pc/ 
[r] 
^{#1} 
\ar@{-} 
@/_1pc/ 
[r] 
_{#2} 
& 
*{\bullet} 
\\} 
} 

\newcommand{\thetagraph}[3]  
{ 
\xymatrix{ *{\bullet} 
\ar@{-} 
@/^1.5pc/ 
[r] 
^{#1} 
\ar@{-} 
@/_1.5pc/ 
[r] 
_{#3} 
\ar@{-} 
[r]^{#2} 
& 
*{\bullet} 
\\} 
}

\newcommand{\unigon}[1]  
{ 
\xymatrix{ *{\bullet} 
\ar@{-} 
[r]^{#1}  
& 
*{\bullet} 
\\} 
} 

\newcommand{\fourtheta}[4]  
{ 
\cxymatrix{ *{\bullet} 
\ar@{-} 
@/^1.5pc/ 
[r] 
^{#1} 
\ar@{-} 
@/_1.5pc/ 
[r] 
_{#4} 
\ar@{-} 
@/^/ 
[r]^{#2} 
\ar@{-} 
@/_/ 
[r]_{#3} 
& 
*{\bullet} 
\\} 
} 

\newcommand{\bigfourtheta}[4]  
{ 
\xymatrixcolsep{2.6pc}
\cxymatrix{*{\bullet} 
\ar@{-} 
@/^1.7pc/ 
[r]
^{#1} 
\ar@{-} 
@/_1.7pc/ 
[r]
_{#4} 
\ar@{-} 
@/^/ 
[r]^{#2} 
\ar@{-} 
@/_/ 
[r]_{#3} 
& 
*{\bullet} 
\\} 
} 

\newcommand{\TenJ}{%
\begin{xy} 
\xygraph{!{<4pc,0pc>:}
  !P5"A"{~><{@{{}{-}*{\bullet}}} ~>>{_{j_{\xypolynode}}}}
  "A1" -@-_{j_6} "A3"
  "A2" -@-_{j_7} "A4"
  "A3" -@-_{j_8} "A5"
  "A4" -@-_{j_9} "A1"
  "A5" -@-_{j_{10}} "A2"
}
\end{xy}
}

\newcommand{\TenJv}{%
\begin{xy} 
\xygraph{!{<5pc,0pc>:}
  !P5"A"{~><{@{{}{-}*{\bullet}}} ~>>{_{j_{\xypolynode}(v)}}}
  "A1" -@-_{j_6(v)} "A3"
  "A2" -@-_{j_7(v)} "A4"
  "A3" -@-_{j_8(v)} "A5"
  "A4" -@-_{j_9(v)} "A1"
  "A5" -@-_{j_{10}(v)} "A2"
}
\end{xy}
}

\def\outerlab{\ifcase\xypolynode\or 0 \or 0 \or 0 \or 0 \or 0 \fi}
\def\innerlab#1{\ifcase#1\or 0 \or 0 \or 0 \or 0 \or 0 \fi}

\newcommand{\TenJzero}{%
\begin{xy} 
\xygraph{!{<4pc,0pc>:}
  !P5"A"{~><{@{{}{-}*{\bullet}}} ~>>{_{\outerlab}}}
  "A1" -@-_{\innerlab1} "A3"
  "A2" -@-_{\innerlab2} "A4"
  "A3" -@-_{\innerlab3} "A5"
  "A4" -@-_{\innerlab4} "A1"
  "A5" -@-_{\innerlab5} "A2"
}
\end{xy}
}

\def\outerlabe{\ifcase\xypolynode\or \frac{1}{2} \or \frac{1}{2} \or 
0 \or 0 \or 0 \fi}
\def\innerlabe#1{\ifcase#1\or \frac{1}{2} \or 0 \or 0 \or 0 \or 0 \fi}

\newcommand{\TenJhalf}{%
\begin{xy} 
\xygraph{!{<4pc,0pc>:}
  !P5"A"{~><{@{{}{-}*{\bullet}}} ~>>{_{\outerlabe}}}
  "A1" -@-_{\innerlabe1} "A3"
  "A2" -@-_{\innerlabe2} "A4"
  "A3" -@-_{\innerlabe3} "A5"
  "A4" -@-_{\innerlabe4} "A1"
  "A5" -@-_{\innerlabe5} "A2"
}
\end{xy}
}

\begin{document} 

\title{Spin Foam Models of Riemannian Quantum Gravity}
\author{John C.\ Baez}
\address{Department of Mathematics\\
         University of California\\
         Riverside, California 92521 USA} 
\email{baez@math.ucr.edu}
\author{J.\ Daniel Christensen}
\address{Department of Mathematics\\
         University of Western Ontario\\
         London, ON N6A 5B7 Canada}
\email{jdc@uwo.ca} 
\author{Thomas R.\ Halford}
\email{thalford@ieee.org} 
\author{David C.\ Tsang}
\email{dtsang@physics.ubc.ca} 

\date{July 21, 2002}

\begin{abstract}      
Using numerical calculations, we compare three versions of the 
Barrett--Crane model of 4-dimensional Riemannian quantum gravity.     
In the version with face and edge amplitudes as described by De Pietri,
Freidel, Krasnov, and Rovelli, we show the partition function diverges
very rapidly for many triangulated 4-manifolds.   In the
version with modified face and edge amplitudes due to Perez and Rovelli,
we show the partition function converges so rapidly that the sum is
dominated by spin foams where all the spins labelling faces are zero
except for small, widely separated islands of higher spin.   We also
describe a new version which appears to have a convergent partition
function without drastic spin-zero dominance.  Finally, after a general
discussion of how to extract physics from spin foam models, we discuss
the implications of convergence or divergence of the partition function
for other aspects of a spin foam model.
\end{abstract}

\maketitle

\section{Introduction}

Despite increasing interest in spin foam models of 4-dimensional quantum
gravity \cite{Baez2,Oriti}, most work so far has focused on the formal
properties of these models, rather than the crucial question of whether
they yield reasonable physics at experimentally accessible length
scales.  Apart from the predilections of the researchers working in this
field, there are two main reasons for this.  

First, it is not obvious which computations would settle this issue.
Second, it is difficult to do any sort of computation with these models.
If the discreteness of a typical spin foam occurs at the Planck scale, a
brute-force simulation of a region of space the size of a proton for the
time it takes light to cross this region would require summing over spin
foams having roughly $10^{80}$ vertices.  Of course, it would be
interesting to simulate even a much smaller spin foam.  However, in the
Barrett--Crane model of 4-dimensional quantum gravity we must compute a
quantity called the $10j$ symbol for each spin foam vertex \cite{BC,BC2}.  
In the Lorentzian versions of this theory, no efficient way to compute the
$10j$ symbol is known so far: the only existing methods are Monte Carlo
calculations that sometimes require over $10^{10}$ samples to achieve
reasonable accuracy \cite{BaezC}.  This makes even very small spin foams
difficult to deal with.

The situation is a bit better for Riemannian versions of the 
Barrett--Crane model.  An efficient algorithm has been developed 
which computes the Riemannian $10j$ symbols in $O(j^5)$ time using
$O(j^2)$ memory, where $j$ is the average of the ten spins 
involved~\cite{CE}.  As an example, on a 1GHz Pentium III CPU, this
algorithm takes about 5 milliseconds to compute the $10j$ symbol with
all spins equal to $\frac{5}{2}$, and about 2.5 seconds to compute the
$10j$ symbol with all spins equal to $10$. This makes it feasible to
compare the qualitative behavior of different versions of the
Riemannian Barrett--Crane model by means of computer calculations for 
small spin foams.  As it turns out, the results are dramatic and 
enlightening.

In what follows, we start with a quick review of the existing spin foam
models of Riemannian quantum gravity.  Then we study three versions of
the Barrett--Crane model of 4-dimensional Riemannian quantum gravity. 
In all three versions the spin foam vertex amplitudes are given by the
Riemannian $10j$ symbols; they differ only in their formulas for edge
and face amplitudes.   In Section \ref{BC} we show that in the model
due to De Pietri, Freidel, Krasnov and Rovelli \cite{DFKR}, the
partition function diverges very rapidly for the simplest triangulation
of the 4-sphere, and probably for many other triangulated 4-manifolds as
well.  In Section \ref{PR} we turn to the model due to Perez and
Rovelli \cite{PerezRovelli}.  Here it is already known that the
partition function converges for any nondegenerate triangulation of any
compact 4-manifold \cite{Perez,Perez2}.  We show that in fact the partition
function converges so fast that the sum over spin foams is dominated
by those where almost all the spins labelling faces are zero.  In
Section \ref{new} we describe a new model with intermediate behavior.
This model seems to have a convergent partition function without 
drastic spin-zero dominance.  In Section \ref{implications} we discuss the 
implications of our results.  

\section{Review} \label{review}

Spin foam models are an attempt to describe the geometry of spacetime in
a way that takes quantum theory into account from the very start.  A
spin foam is a 2-dimensional analogue of a Feynman diagram.  Abstractly,
a Feynman diagram can be thought of as a graph with edges labelled by
group representations and vertices labelled by intertwining operators. 
Similarly, a spin foam is a 2-dimensional cell complex with polygonal
faces labelled by representations and edges labelled by intertwining
operators.  Like Feynman diagrams, spin foams serve as a basis of
`quantum histories': the actual time evolution of the system is
described by a linear combination of these quantum histories,  weighted
by certain amplitudes.  Feynman diagrams are 1-dimensional because they
describe quantum histories of collections of point particles; spin foams
are 2-dimensional because in loop quantum gravity, the gravitational
field is described not in terms of point particles but 1-dimensional
`spin networks'.  

An ordinary quantum field theory provides a recipe for computing the
amplitude for any Feynman diagram in terms of amplitudes for edges and
vertices.  Similarly, a spin foam model consists of a recipe to compute
an amplitude for any spin foam as a product of face, edge and vertex
amplitudes. The partition function in a spin foam model is computed as a
sum or integral of these spin foam amplitudes.  Using suitably weighted
sums and normalizing by dividing by the partition function,  one can
also compute expectation values of observables.

A number of spin foam models have been developed for both Lorentzian and
Riemannian quantum gravity.  By `Lorentzian quantum gravity', we
mean any quantum theory whose partition function is, at least morally
speaking, given by 
\[     \int e^{iS} , \] 
where $S$ is the Einstein--Hilbert action for a Lorentzian metric on
spacetime, or some closely related action.   If all goes well, a theory
of this sort should reduce in a suitable limit to the classical Einstein
equations for Lorentzian metrics.  `Riemannian quantum gravity'  is the
same sort of thing, but for Riemannian metrics.  It is important not to
confuse Riemannian quantum gravity with `Euclidean quantum gravity', which 
also uses the Einstein--Hilbert action for a Riemannian metric, but where 
the partition function is given by
\[    \int e^{-S}  .\]  
With the help of analytic continuation to imaginary times, Euclidean
quantum gravity is a widely used (though controversial) tool for studying
Lorentzian quantum gravity.  Riemannian quantum gravity is a wholly
different theory, which at best will reduce in some limit to the
classical Einstein equations for Riemannian metrics.  Thus the large
body of conventional wisdom about Euclidean quantum gravity may not
apply to spin foam models of Riemannian quantum gravity --- only further
work can decide this.

Riemannian quantum gravity seems to have limited relevance to real-world
physics. Nonetheless, spin foam models of Riemannian quantum gravity
have proved to be a useful warmup for work on spin foam models of
Lorentzian quantum gravity.  The Riemannian models are simpler because
the rotation group is compact, unlike the Lorentz group.  For a compact
group, the irreducible unitary representations are finite-dimensional
and indexed by discrete rather than continuous parameters.  This means
that in Riemannian spin foam models there is no difficulty showing the
convergence of a single spin foam amplitude, and the partition function
is computed as a sum rather than an integral over spin foams.

In retrospect, the very first spin foam model was the Ponzano--Regge
model of 3-dimensional Riemannian quantum gravity \cite{PR}.   In this
model, one triangulates a given 3-manifold and expresses the partition
function $\int e^{iS}$ as a sum over spin foams lying in the dual
2-skeleton of the triangulation.   The gauge group in this theory is
the double cover $\Spin(3) = \SU(2)$ of the 3d rotation group.   A
heuristic argument suggested that the result was actually independent of
the triangulation. In fact the sum diverges, and contrary to Ponzano and
Regge's original expectations, the naive way of regularizing it does not
give triangulation independent results.  

Much later, Turaev and Viro \cite{TV} discovered that one could
regularize the Ponzano-Regge model by replacing $\SU(2)$ with the
corresponding quantum group, $\SU_q(2)$.  In this $q$-deformed model the
partition function converges for any compact 3-manifold.  Even better,
it turns out to be triangulation independent.  By now it is clear that
this partition function is that of 3-dimensional Riemannian quantum
gravity with a positive cosmological constant; the deformation parameter
$q$ is related to the cosmological constant by a simple formula~\cite{AW,MT}.

In 1997, one week before the concept of spin foam was formalized
\cite{Baez}, Barrett and Crane proposed a spin foam model of
4-dimensional Riemannian quantum gravity \cite{BC}.  Here the partition
function is computed as a sum over spin foams lying in the dual
2-skeleton of a triangulated 4-manifold.  The spin foam faces are all
labelled by `balanced' representations of $\Spin(4) = \SU(2) \times
\SU(2)$, that is, those of the form $j \tensor j$.  The edges are all
labelled by a specific intertwiner called the Barrett--Crane
intertwiner.  The idea behind these choices was to express Riemannian
quantum gravity as a constrained version of the spin foam model for
topological gravity due to Turaev, Ooguri, Crane and Yetter
\cite{CY,Ooguri,Turaev}.   This idea also motivated using the $10j$
symbols for the vertex amplitudes.   

Barrett and Crane wisely refrained from giving formulas for edge and
face amplitudes, which later turned out to be the most controversial
aspect of the whole theory.   Unfortunately, without these, their model
was incomplete.  In particular, it was impossible to say whether or not
the partition function converges.  They did note that one can $q$-deform
their model, obtaining a model based on the quantum group $\SU_q(2)
\times \SU_{\overline{q}}(2)$.  In this $q$-deformed version the
partition function becomes a finite sum, so it converges regardless of
the choice of edge and face amplitudes.  However, nobody has been able
to find a nontrivial choice of edge and face amplitudes that makes the
result triangulation independent.   By analogy with the 3-dimensional
case, it is widely expected that the $q$-deformed Barrett--Crane model
is related to 4-dimensional Riemannian quantum gravity with positive
cosmological constant.   Since this is not a topological field theory,
there is actually no reason to expect a triangulation-independent partition
function.

Further progress was made by De Pietri, Freidel, Krasnov and Rovelli
\cite{DFKR}, who showed that the Barrett--Crane model naturally arises
from a quantum field theory on a product of 4 copies of the 3-sphere,
thought of as a homogeneous space of the group $\SU(2) \times \SU(2)$.
Feynman diagrams in this `group field theory' correspond precisely to
the spin foams appearing in the Barrett-Crane model, and the vertex
amplitudes are the same as well.  There are two important differences,
however.  First, the group field theory approach gives specific formulas
for edge and face amplitudes!  Second, instead of summing over spin
foams lying in the dual 2-skeleton of a fixed triangulation of a fixed
4-manifold, one computes the partition function by summing over spin
foams lying in the dual 2-skeleta of {\it all} triangulations of {\it
all} compact 4-manifolds --- and even a more general class of
well-behaved `pseudomanifolds', namely spaces made by gluing finitely
many 4-simplices together pairwise along their tetrahedral faces.  In
short, the DFKR approach naturally extends the Barrett--Crane model to
incorporate a sum over triangulations and even a sum over topologies.
This sidesteps the awkward need for an arbitrary choice of triangulation, 
but makes the convergence of the partition function even less likely.

Later, Perez and Rovelli \cite{PerezRovelli} modified the DFKR proposal,
describing a group field theory that corresponds to a version of the
Barrett--Crane model with modified edge and face amplitudes.  Their 
goal was to eliminate divergences from the model, and they made
substantial progress: Perez \cite{Perez,Perez2} was able to prove that
in this modified model, the sum of spin foam amplitudes converges if we
restrict to spin foams lying in the dual 2-skeleton of a given
well-behaved pseudomanifold, so long as each triangle of this 
pseudomanifold lies in at least three 4-simplices.   The issue of
pseudomanifolds not satisfying this condition remains a challenge,
as does the sum over pseudomanifolds.

As we shall see in Section \ref{implications}, it is not completely
obvious that one needs a convergent partition function for a
well-behaved spin foam model.  After all, for physics we need to
compute, not the partition function, but expectation values of
observables.  Only after we can compute these can we tackle the
important question of whether a given spin foam model reduces to
general relativity (possibly coupled to matter) in the large-scale
limit.  Nonetheless, it appears that convergence or divergence of the
partition function is closely tied to other important qualitative
features of a spin foam model.  This makes it worthwhile to study the
convergence issue.  In the next three sections, we do this for three
versions of the Riemannian Barrett--Crane model: the DFKR version, the
Perez--Rovelli version, and a new version.  We only consider the
convergence issue for one pseudomanifold at a time, not the sum over
pseudomanifolds.

\section{The De Pietri--Freidel--Krasnov--Rovelli model} \label{BC}

We begin by recalling the general idea of the Riemannian Barrett--Crane
model.  As mentioned already, this model can be defined for any
simplicial complex formed by taking a finite set of 4-simplices and
attaching distinct ones pairwise along their tetrahedral faces 
until all faces are paired.  In what follows we shall restrict 
attention to manifolds, but only to simplify the terminology; 
everything generalizes painlessly to these well-behaved pseudomanifolds.

Let $M$ be a triangulated compact 4-manifold and let $\Delta_n$ be the
set of $n$-simplices in the triangulation.  By definition, 
4-simplices, 3-simplices and 2-simplices correspond to vertices, edges
and faces of the dual 2-skeleton, respectively.  In all versions of the
Riemannian Barrett--Crane model, a spin foam $F$ simply amounts to a
labelling of each face $f \in \Delta_2$ by a spin $j(f) \in \{0,
\frac{1}{2}, 1, \dots\}$.   Note that there are four faces incident to
each edge in the dual 2-skeleton.  We require that the spins $j_1,
\dots, j_4$ labelling the faces incident to any edge be `admissible',
meaning that there exists a nonzero $\SU(2)$ intertwining operator $f
\maps j_1 \tensor j_2 \to j_3 \tensor j_4$.

In all versions of the Riemannian Barrett--Crane model, the 
amplitude of a spin foam is computed by a formula of this sort: 
\begin{equation}
     Z(F) = \prod_{f \in \Delta_2} A(f) 
            \prod_{e \in \Delta_3} A(e) 
            \prod_{v \in \Delta_4} A(v) 
\label{spin.foam.amplitude}
\end{equation}
and the partition function is given by
\begin{equation}
    Z(M) = \sum_F Z(F)  .
\label{partition.function}
\end{equation}
Here the complex numbers $A(f), A(e)$ and $A(v)$ are called face, edge
and vertex amplitudes, respectively.  Each face amplitude is computed
only using the spin $j(f)$ labelling that face.  Each edge amplitude is
computed using the spins labelling the 4 faces incident to that edge; we
call these spins $j_1(e), \dots, j_4(e)$.  Finally, each vertex
amplitude is computed using the spins labelling the 10 faces incident to
that vertex; we call these spins $j_1(v), \dots, j_{10}(v)$.  

To give formulas for these amplitudes we use the standard graphical
notation for $\SU(2)$ spin networks \cite{CFS,KL}.   Normalization
issues are crucial here.  We call a triple of spins $j_1,j_2,j_3$
`admissible' if there exists a nonzero intertwining operator 
$f \maps j_1 \tensor j_2 \to j_3$; this happens precisely when these
spins satisfy the triangle inequality and sum to an integer.  
Given an admissible triple of spins, we normalize the canonical 
intertwining operator $f \maps j_1 \tensor j_2 \to j_3$ so that
\[
\thetagraph{j_1}{j_2}{j_3} = 1.
\]
As this normalization sometimes requires dividing by the square root of
a negative number, it introduces a potential sign ambiguity.  Luckily,
in our calculations trivalent vertices always come in matching pairs, so
these signs cancel.  

Actually, in what follows almost all our diagrams will be balanced spin
networks \cite{BC,Yetter}.  In such a spin network, labelling an edge by
the spin $j$ really means that it is labelled by the irreducible
representation $j \tensor j$ of the group $\Spin(4) = \SU(2) \times
\SU(2)$.  Such representations are called `balanced'.  Also, in a
balanced spin network, an unlabelled 4-valent vertex is really labelled
by the Barrett-Crane intertwiner.  This is defined in terms of $\SU(2)$
spin networks by:
\begin{equation}
\FourX{j_1}{j_2}{j_3}{j_4} = \sum_k \, (-1)^{2k} (2k+1) \;
\DoubleY{j_1}{j_2}{j_3}{j_4}{k} \tensor \DoubleY{j_1}{j_2}{j_3}{j_4}{k}
, \label{BC.intertwiner} 
\end{equation} 
where we sum over spins $k$ such that the triples $j_1,j_2,k$ and
$j_3,j_4,k$ are both admissible.   

In terms of balanced spin networks, the face, edge and vertex amplitudes 
of the DFKR model are given as follows:
\begin{equation}
\begin{array}{lcc} 
A(f) &=& \monogon{j(f)} \\ 
     &&                                            \\
A(e) &=& \frac{\displaystyle 1}
              {\ \ \bigfourtheta{j_1(e)}{j_2(e)}{j_3(e)}{j_4(e)}\ \ } \\ 
     &&                                             \\
A(v) &=& \TenJv .
\end{array}
\label{BC.amplitudes}
\end{equation}
Here the intertwiner in the first spin network is just the
identity operator, so the 
face amplitude $A(f)$ is just the dimension of the representation
$j(f) \tensor j(f)$, that is, $(2j(f) + 1)^2$.  The `$4j$ symbol'
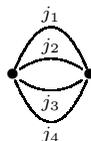
\[    \fourtheta{j_1}{j_2}{j_3}{j_4} \]
equals the dimension of the space of $\SU(2)$ intertwining operators
\[  
    f \maps j_1 \tensor j_2
       \to  j_3 \tensor j_4 .\]
This in turn equals the number of spins $k$ such that 
the triples $j_1,j_2,k$ and $j_3,j_4,k$ are both admissible.
Finally, the vertex amplitude $A(v)$ is called the `$10j$ symbol'.
There is no simple formula for this, so to compute it we shall need
the algorithm developed by Christensen and Egan \cite{CE}.  

At this point some comments might be helpful.  The above formulas were
first derived by De Pietri, Freidel, Krasnov and Rovelli \cite{DFKR}
using the group field theory approach.  However, they also arise
naturally from the idea that the Barrett--Crane model is a constrained
version of the $\SU(2) \times \SU(2)$ Turaev--Ooguri--Crane--Yetter
model.  In the latter model one works with spin foams where faces are
labelled by arbitrary irreducible representations of $\SU(2) \times
\SU(2)$ and edges are labelled by arbitrary intertwiners {\it chosen
from an orthonormal basis}.  Here one uses the fact that intertwiners $f
\maps H \to H'$ between finite-dimensional unitary group representations
naturally form a Hilbert space with
\[      \langle f, g \rangle = \tr(f^* g)  .\]
To get the above version of the Barrett--Crane model, one restricts the 
Turaev--Ooguri--Crane--Yetter formulas to spin foams where faces are
labelled by balanced intertwiners and edges are labelled by the
{\it normalized} Barrett--Crane intertwiner.  The Barrett--Crane 
intertwiner in equation (\ref{BC.intertwiner}) is not normalized; 
instead, its inner product with itself is 
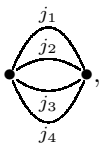
\[    \fourtheta{j_1}{j_2}{j_3}{j_4}   , \]
so to normalize it we must divide by the square root of this quantity.
However, since each Barrett--Crane intertwiner in the $10j$ symbols appears
twice in the formula for the $Z(F)$ --- once for each of the two 4-simplices
incident to a given 3-simplex --- we obtain a factor of
\[ \frac{\displaystyle 1}
        {\ \ \bigfourtheta{j_1(e)}{j_2(e)}{j_3(e)}{j_4(e)}\ \ } , \]
which gives the edge amplitude $A(e)$.  

We can study the convergence of the partition function
(\ref{partition.function}) by imposing a cutoff on the spins labelling
spin foam faces.   Since these spins determine the areas of the
corresponding 2-simplices in the triangulation of the spacetime
manifold, we can think of this as a sort of `infrared cutoff' which
rules out large areas.  Let us write $|F|$ for the maximum of the
spins labelling the faces of the spin foam $F$.  Imposing a spin
cutoff $|F| \le J$, the partition function becomes a finite sum
\begin{equation}
Z_J(M) = \sum_{|F| \le J} Z(F) .
\label{cutoff.partition.function}
\end{equation}
For a simple but interesting example, we can take $M$ to be a 4-sphere
triangulated as the boundary of a 5-simplex.  In Table 1 we show the
results of computing $Z_J(M)$ in this case for various low values of 
$J$.   

\vskip 1em
{\vbox{   
\begin{center}   
{\small
\setlength{\extrarowheight}{1.5pt}
\setlength{\tabcolsep}{2.2pt}
\begin{tabular}{|c|D{X}{\,\cdot\,}{4}|}            \hline
$J$     & \multicolumn{1}{c|}{$Z_J(M)$}   \\ \hline
$0$     &                 1.000 X 10^{0}    \\ \hline
$1/2$   &                 3.722 X 10^{5}     \\ \hline
$1$     &                 7.812 X 10^{9}     \\ \hline
$3/2$   &                 2.128 X 10^{13}     \\ \hline
$2$     &                 1.345 X 10^{16}     \\ \hline
\end{tabular}}
\end{center}   
\vskip 1em 
\centerline{Table 1: $S^4$ partition function --- DFKR model with 
spin cutoff $J$}
}}   
\vskip 1em   

It seems that $Z_J(M)$ grows at a spectacular rate as $J$ increases.  We
can begin to understand this by estimating the face, edge and vertex
amplitudes in equation (\ref{BC.amplitudes}). In the limit of large
spins, the face amplitudes clearly grow as $O(j^2)$ where $j$ is the
spin labelling the face in question.   For the edge amplitudes, we can
use the fact that
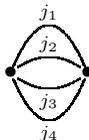
\[    \fourtheta{j_1}{j_2}{j_3}{j_4} \]
equals the number of spins $k$ such that both $j_1, j_2, k$ and $j_3,
j_4, k$ are admissible triples.  In general, if $j_1, \dots, j_4$ are
admissible and of order $j$, the number of such spins $k$ is also of
order $j$, so the edge amplitudes grow as $O(j)$ when all spins are
rescaled by the same factor.  The only exception occurs when
$j_1,\dots,j_4$ lie at the `border of admissibility', that is, when 
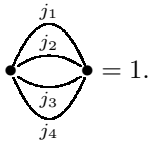
\[    \fourtheta{j_1}{j_2}{j_3}{j_4} = 1 .\]
In this case the $4j$ symbol remains equal to one as all four spins
are rescaled.   

The asymptotic behavior of the vertex amplitude is much more subtle.
Starting from the integral formula for the $10j$ symbols and doing a
stationary phase approximation, Barrett and Williams \cite{BW} computed
the asymptotics of the $10j$ symbols as all ten spins are multiplied by
some factor $j$ which approaches infinity.  This calculation yields a
factor of $j^{-9/2}$ times an oscillating function of $j$.
Unfortunately, computer calculations show a different rate of decay and
no significant oscillatory behavior \cite{BCE}.  It now seems clear that
the stationary phase points do not dominate the integral.  In general,
it appears that the $10j$ symbols decay as $O(j^{-2})$ as all spins are
rescaled by the same factor.  The only exception occurs when the four
spins labelling edges incident to some vertex lie at the border of
admissibility; then the $10j$ symbols decay more rapidly.  The more
vertices lie at the border of admissibility, the more rapid the decay.

The triangulation of the 4-sphere as the boundary of a 5-simplex gives
spin foams with 20 faces, 15 edges and 6 vertices.   Thus, for a spin
foam $F$ with all faces labelled by spins of order $j$, with no 
vertices lying on the border of admissibility, the amplitude is
\[  Z(F) = O(j^{2 \cdot 20} j^{-1 \cdot 15} j^{-2 \cdot 6}) 
         = O(j^{13})  . \]
Together with the fact that $Z(F)$ is always nonnegative \cite{BaezC},
so that no cancellations are possible, this is already enough to show
that $Z_J(M) \to +\infty$ as $J \to +\infty$.  In fact, just by summing 
over spin foams where all faces are labelled by the same spin, we
already see that $Z_J(M)$ must tend to infinity at a rate no slower than
$J^{14}$.  However, this is a drastic underestimate.  In fact, doing a
least squares fit to a log-log plot of the above tabulated values of
$Z_J(M)$, we estimate that $Z_J(M)$ grows as approximately $J^{23}$.

  From these considerations it is clear that the partition function in the
DFKR model will diverge, not just in this example, but for many
triangulated 4-manifolds.  Since the divergence is mainly due to rapid
growth of the face amplitudes with increasing spin, it seems the partition
function only has a chance of converging if there are few spin foam faces
compared to spin foam edges and vertices.  Spin foams of this sort come
 from triangulations where there are few triangles compared to tetrahedra
and 4-simplices.  Since the number of tetrahedra in a nondegenerate
triangulation is always $\frac{5}{2}$ times the number of 4-simplices,
this occurs when the average number of 4-simplices meeting along each
triangle is high.

As explained more carefully in Section \ref{implications}, we can
try to extract physical information from spin foam models by computing
expectation values of observables.  These are weighted averages of
functions assigning to each spin foam a real number, where the weight
associated to each spin foam is its amplitude.  The Metropolis
algorithm \cite{Metropolis} is a powerful tool for numerically
computing weighted averages, but only when the weights are
nonnegative.  Quantum amplitudes are usually complex, so the
Metropolis algorithm is normally applicable to quantum theory problems
only after one has converted them into statistical mechanics problems
via Wick rotation.  Luckily, the amplitude for a spin foam in the DFKR
model is always nonnegative!  In fact, this is true for all the
versions of the Riemannian Barrett--Crane model that we consider here
\cite{BaezC}.  This allows us to apply the Metropolis algorithm
without becoming enmeshed in the subtleties of Wick rotation.  As we
shall see, this algorithm provides great insight into which spin foams
dominate the partition function.

For readers unfamiliar with the Metropolis algorithm, let us briefly
describe how it works in general before turning to our application
here.  This algorithm is a random walk technique for sampling
configurations from some finite set $\Lambda$ with desired relative
frequencies given by some function $p \maps \Lambda \to [0,\infty)$.
To use the algorithm one must first choose a finite set of `moves'
$f_j \maps \Lambda \to \Lambda$.  Starting at an arbitrary
configuration $x_{0} \in \Lambda$, one then generates a random
sequence of configurations $x_i$ as follows.  For each $i$, choose a
move $f_{j}$ uniformly at random.  Let $x_{i+1} = f_{j}(x_{i})$ with
probability $p(f_{j}(x_{i}))/p(x_{i})$ and otherwise let $x_{i+1} =
x_{i}$.  (If $p(f_{j}(x_{i}))/p(x_{i}) > 1$, always set $x_{i+1} =
f_{j}(x_{i})$.)  If the moves are chosen appropriately, the
distribution of the entries $x_{i}$ in the sequence will tend to the
desired distribution $p$ as $i$ tends to infinity.

To get the right limiting distribution, the moves must be chosen
`large enough' so that the algorithm is ergodic, i.e., so that any
configuration $x$ with $p(x)>0$ has a nonzero chance of occurring.
However, to get reasonably fast convergence to the limiting distribution,
the moves must be chosen `small enough' so that the algorithm
spends much of its time moving between highly weighted configurations,
rather than spinning its wheels rejecting configurations with small $p$.
Choosing the moves to balance these opposing needs is a bit of an art.
In practice, one determines experimentally whether one's moves work
well.

To use the Metropolis algorithm to compute expectation values of
observables in some version of the Riemannian Barrett--Crane model, 
we want to sample spin foams with relative frequencies given by their
amplitudes.  To do this we take $\Lambda$ to be some finite set of
spin foams and let $p \maps \Lambda \to [0,\infty)$ assign to each
spin foam its amplitude.  We also need to choose a set of moves for
going from one spin foam to another.  For our example of the 4-sphere
triangulated as the boundary of a 5-simplex, we chose to use moves
that consist of picking a tetrahedron in the dual 2-skeleton and
adding or subtracting $\frac{1}{2}$ to each of the spins labelling the
four faces of this tetrahedron.  The $10j$ symbols vanish unless the
sum of the spins labelling faces incident to each edge is an integer.
Our moves preserve this constraint.  If subtracting $\frac{1}{2}$ from
a spin would make it negative, we leave all the spins unchanged, but
still count this process as a move.  In our experiments, this
collection of moves produces very fast convergence of the Metropolis
algorithm, and the predicted answer is accurate in all cases in which
we have been able to compute the exact answer by other means.

It is interesting to compare the behavior of this algorithm for various
versions of the Barrett--Crane model.  Unfortunately, in the DFKR version, 
the divergence of the partition function means that the random walk will
drift toward spin foams with ever larger spins, since these have the 
largest amplitudes and there are many of them.  Table 2 shows a
small portion of a typical run of the Metropolis algorithm for this 
version, with a spin cutoff of $J = \frac{5}{2}$.   The first column 
is the iteration number.  In steps that are not shown, the program 
stayed at the same labelling.  The second column displays the twenty 
spins labelling faces, each multiplied by two.  The third column shows 
the amplitude of the corresponding spin foam.  One can see that the 
sum over spin foams is dominated by those with many spins close to the 
cutoff.  

\vskip 1em

{\vbox{   
\begin{center}   
{\small
\setlength{\extrarowheight}{1.5pt}
\setlength{\tabcolsep}{2.2pt}
\begin{tabular}{|>{$}c<{$}|>{$}c<{$}|D{X}{\,\cdot\,}{3}|}    \hline
\text{iteration} & F          & \multicolumn{1}{c|}{$Z(F)$}\\ \hline
335291 & 34234252435354544545 & 4.4 X 10^{9} \\ \hline
335296 & 34234152435454545555 & 3.1 X 10^{9} \\ \hline
335302 & 34244142345454545555 & 1.9 X 10^{9} \\ \hline
335303 & 34344043335454545555 & 5.6 X 10^{8} \\ \hline
335304 & 34444043335555555555 & 1.0 X 10^{9} \\ \hline
335310 & 24443133335555555555 & 3.4 X 10^{9} \\ \hline
335312 & 23444132235555555555 & 2.0 X 10^{9} \\ \hline
335320 & 13443242235555555555 & 1.1 X 10^{9} \\ \hline
335321 & 04533242235555555555 & 2.5 X 10^{8} \\ \hline
335323 & 04543252345555555555 & 4.6 X 10^{8} \\ \hline
335324 & 04544252344555554455 & 3.9 X 10^{8} \\ \hline
335327 & 05545251244555554455 & 9.9 X 10^{7} \\ \hline
335328 & 05445150254555554455 & 1.5 X 10^{7} \\ \hline
335351 & 05445250254455555445 & 1.4 X 10^{7} \\ \hline
\end{tabular}} 
\end{center}   
\vskip 1em 
\centerline{Table 2: sample Metropolis labellings --- DFKR model
with spin cutoff $\frac{5}{2}$}
}}   
\vskip 3em   

\section{The Perez--Rovelli Model} \label{PR}

In the Perez--Rovelli model, the face, edge and vertex amplitudes are as
follows:
\begin{equation}
\begin{array}{lcc} 
A(f) &=& \monogon{j(f)}   \\ 
     &&                                                \\
A(e) &=& \frac{\bigfourtheta{j_1(e)}{j_2(e)}{j_3(e)}{j_4(e)} }
              { \monogon{j_1(e)}\; \monogon{j_2(e)} \;
                \monogon{j_3(e)} \; \monogon{j_4(e)}  }    \\
     &&                                                \\
A(v) &=& \TenJv .
\end{array}
\label{PR.amplitudes}
\end{equation}
Here again a comment is in order: the original papers by Perez and
Rovelli \cite{Perez,PerezRovelli} give a different formula for the edge
amplitudes, but that formula does not really follow from their group
field theory.  Above we use the corrected formula which appears in a
forthcoming review article by Perez \cite{Perez2}; we have carefully
translated from his normalization conventions to our own.

With these formulas, Perez has shown that the partition function
converges for any well-behaved pseudomanifold satisfying the condition
that each triangle lies in at least three 4-simplices.  This includes
the triangulation of $S^4$ as the boundary of a 5-simplex.  Nonetheless
it is illuminating to compute the cutoff partition function $Z_J(M)$ in
this example with various choices of the spin cutoff.  The results
appear in Table 3.

\vskip 1em

{\vbox{   
\begin{center}   
{\small
\setlength{\extrarowheight}{1.5pt}
\setlength{\tabcolsep}{2.2pt}
\begin{tabular}{|c|c|}    \hline
$J$ & $Z_J(M)$                                      \\ \hline
0 &      1.000000000000                 \\ \hline
$1/2$ &  1.000014319178                  \\ \hline
1 &      1.000014323656                  \\ \hline
$3/2$ &  1.000014323670                 \\ \hline
2 &      1.000014323670                 \\ \hline
\end{tabular}} 
\end{center}   
\vskip 1em 
\centerline{Table 3: $S^4$ partition function ---
Perez--Rovelli model with spin cutoff $J$}
}}   
\vskip 1em   

This time it appears that the {\it convergence} is spectacularly rapid.
But again, this is easy to understand: it is mainly due to the
denominator of the edge amplitude in formula (\ref{PR.amplitudes}).
Since each triangle in this triangulation of the 4-sphere is the 
face of 3 tetrahedra, the factors of 
\[       \monogon{j(f)} = (2j(f) + 1)^2 \]
appearing in the edge and face amplitudes combine to give
\[   Z_J(M) = \sum_{|F| \le J}  \;
            \prod_{f \in \Delta_2} (2j(f) + 1)^{-4}
            \prod_{e \in \Delta_3}\bigfourtheta{j_1(e)}{j_2(e)}{j_3(e)}{j_4(e)}
            \prod_{v \in \Delta_4} A(v). \]
The factors of $(2j(f) + 1)^{-4}$ strongly suppress faces labelled
by nonzero spins.   The $10j$ symbols also tend to suppress nonzero
spins.  While the $4j$ symbols grow with increasing spin, they
do so too slowly to make much of a difference.

In particular, $Z_0(M) = 1$ because there is one spin foam with all
faces labelled by spin $0$, and every balanced spin network with all
edges labelled by spin $0$ evaluates to $1$.  In computing
$Z_{1/2}(M)$, we must also consider spin foams where some faces
are labelled by spin $\frac{1}{2}$.  At each spin foam edge, if one of
the incident faces is labelled by a nonzero spin, then at least one
other must be as well, or else the Barrett-Crane intertwiner  there will
vanish.  This is a powerful constraint.  For the 4-sphere triangulated
as the boundary of a 5-simplex, it implies that if there is one spin
foam face labelled by a nonzero spin, then there must be at least four.
When four of the faces are labelled by spin $\frac{1}{2}$, the factors
of $(2j(f) + 1)^{-4}$ multiply to give $2^{-16}$.  Spin foams with more 
nonzero spins, or spins greater than $\frac{1}{2}$, will be suppressed 
even further.  

In fact, it is instructive to work out by hand the contribution to the
partition function given by spin foams with four spin foam faces
labelled by $\frac{1}{2}$ and the rest zero.  To give a nonzero
result, the four faces must form a tetrahedron in the dual 2-skeleton.
This results in a triangular spin-$\frac{1}{2}$ loop in four of
the $10j$ symbols.  Since
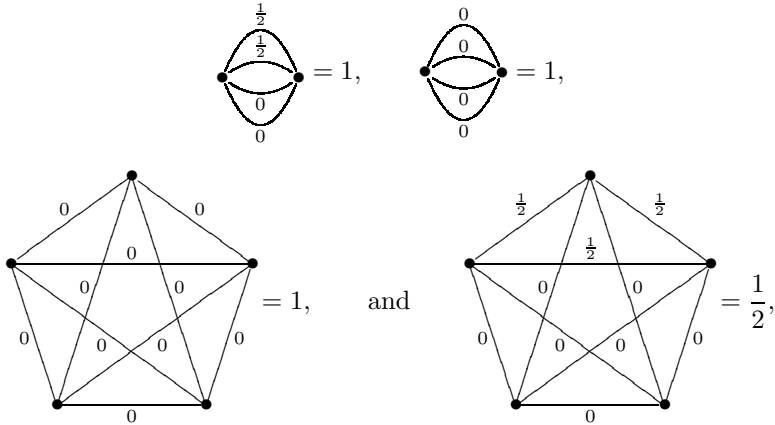
\[   \fourtheta{\frac{1}{2}}{\frac{1}{2}}{0}{0} = 1 , \qquad
     \fourtheta{0}{0}{0}{0} = 1 , \]
\[   \TenJzero = 1 , \qquad \text{and} \qquad
     \TenJhalf  = \frac{1}{2}  , \]
each spin foam of this sort contributes an amplitude of $2^{-16} \cdot
(\frac{1}{2})^4 = 2^{-20}$.    There are $\binom{6}{4} = 15$ spin foams
of this sort, so their total contribution to the partition function is
$15 \cdot 2^{-20} \cong .0000143051$.  Glancing at Table 3, we see
that together with the spin foam having all faces labelled by spin zero,
this accounts for the first seven decimal places of the partition
function.

As a result, when we use the Metropolis algorithm to randomly walk
through spin foams in this example, it focuses attention on  spin foams
where almost all spins are zero.  Table 4 shows a complete run of the
algorithm with a cutoff of $5/2$ and 5 million iterations.  As in Table
2, the first column is the iteration number.  In steps that are not
shown, the program stayed at the same labelling.  The second column
displays the twenty spins labelling faces, each multiplied by two.  The
third column shows the amplitude of the corresponding spin foam.  One
can see that that after 256 steps the initial spin foam has randomly
walked to the spin foam with all faces labelled by by spin zero.  Except
for a brief foray to a spin foam with four spin-$\frac{1}{2}$ faces
between moves 611050 and moves 611136, the algorithm spends all the rest
of its time at the spin foam with all spin-zero faces.

\vskip 1em

{\vbox{   
\begin{center}   
{\small
\setlength{\extrarowheight}{1.5pt}
\setlength{\tabcolsep}{2.2pt}
\begin{tabular}{|>{$}r<{$}|>{$}c<{$}|D{X}{\,\cdot\,}{5}|}    \hline
\text{iteration} & F          & \multicolumn{1}{c|}{$Z(F)$}\\ \hline
     0 & 00110000010111110011 & 1.421 X 10^{-14}      \\    \hline
     1 & 00100000010101020011 & 2.874 X 10^{-12}      \\    \hline
   256 & 00000000010000010011 & 9.537 X 10^{-7}       \\    \hline
   458 & 00000000000000000000 & 1.000 X 10^0          \\    \hline
611050 & 00100101010000000000 & 9.537 X 10^{-7}       \\    \hline
611136 & 00000000000000000000 & 1.000 X 10^0           \\    \hline
\end{tabular}} 
\end{center}   
\vskip 1em 
\centerline{Table 4: sample Metropolis labellings --- Perez-Rovelli model
with spin cutoff $\frac{5}{2}$}
}}   
\vskip 1em   

Generalizing from this example, we can easily guess the behavior of the
Perez--Rovelli model on an arbitrary triangulated 4-manifold.   The
partition function will be dominated by spin foams having mostly
spin-zero faces, and a low density of small islands of faces with
higher spin.    In fact, we can use the `dilute gas' approximation 
\cite{Baez3} to estimate the density of a particular sort of island in
the spin foams that would most often be sampled by the Metropolis
algorithm.  For example, if we consider tetrahedra in the dual
2-skeleton, most of them will have all faces labelled by spin $0$. 
About one in $2^{20}$ will have all four faces labelled by spin
$\frac{1}{2}$, and an even smaller fraction will have faces labelled by
higher spins.  We discuss the implications of this `spin-zero
dominance' in Section \ref{implications}.   

\section{A New Model} \label{new}

Since the partition function of the DFKR model diverges rapidly, while
that of the Perez--Rovelli model converges so rapidly that the sum is
dominated by spin foams with almost all faces labelled with spin zero,
it seems worthwhile to seek a model with intermediate behavior.  It
would be nice to derive this model from a group field theory.  However,
one can also take an exploratory attitude and simply seek face, edge and
vertex amplitudes that give partition functions `near the brink of
convergence', but on the convergent side.  

Comparing various candidates, we found this model to be the most
promising:
\begin{equation}
\begin{array}{lcc} 
A(f) &=& 1   \\ 
     &&                                                \\
A(e) &=&  \frac{\displaystyle 1}
      {\ \ \bigfourtheta{j_1(e)}{j_2(e)}{j_3(e)}{j_4(e)}\ \ }  \\
     &&                                                \\
A(v) &=& \TenJv .
\end{array}
\label{new.amplitudes}
\end{equation}
The first thing to note is this model's simplicity.  As in the DFKR
model, the edge amplitudes arise naturally from normalizing the
Barrett--Crane intertwiners in the $10j$ symbol.  But unlike the DFKR
model, this new model has trivial face amplitudes.  Thus the only real
ingredient of this model is the $10j$ symbol built from normalized
Barrett--Crane intertwiners.  The absence of loops 
\[   \monogon{j}  \]
in the above formulas is the main reason the model lies near the brink
of convergence.  These loops grow rapidly as a function of $j$, so they
tend to make the partition function diverge or converge very quickly,
depending on whether more of them appear in the numerator or denominator
in the partition function.

Table 5 shows the partition function of our new model for the
triangulation of $S^4$ as the boundary of a 5-simplex, as a function of
the spin cutoff $J$.  Though the partition function appears to be
converging, it is hard to be sure from this limited data.
Unfortunately, the calculation of $Z_{5/2}(M)$ already involved a sum
over approximately $3.6$ trillion spin foams.  (There are a total of
$6^{20}$ ways to label all twenty faces with spins from $0$ to
$\frac{5}{2}$, but of these, only $6^{20}/2^{10} \cong 3.6 \cdot
10^{12}$ satisfy the constraint that the spins labelling faces incident
to any edge sum to an integer; only these can give a nonzero result,
so we only summed over these.)  This calculation occupied
28 CPUs for 23 hours.  Going further with this brute-force approach
would require much longer.

\vskip 1em

{\vbox{   
\begin{center}   
{\small
\setlength{\extrarowheight}{1.5pt}
\setlength{\tabcolsep}{2.2pt}
\begin{tabular}{|c|c|}    \hline
$J$ & $Z_J(M)$                                      \\ \hline
0 &       1.000000000000   \\ \hline
$1/2$ &   2.342658607645   \\ \hline
1 &       3.378038633798   \\ \hline
$3/2$ &   3.966290480574   \\ \hline
2 &       4.293589340364   \\ \hline
$5/2$ &   4.480621474940   \\ \hline
\end{tabular}} 
\end{center}   
\vskip 1em 
\centerline{Table 5: $S^4$ partition function --- the new model with
spin cutoff $J$}
}}   
\vskip 1em   

In Section \ref{implications} we describe an indirect method which gives
stronger evidence that the partition converges for this triangulation of
$S^4$.  Of course, one would really like a mathematical proof that the
partition function converges --- and not just in this case, but more
generally.  Numerical calculations show that it
diverges for the well-behaved pseudomanifold formed by taking two 
4-simplices and gluing them together along all their tetrahedral faces.  
However, this leaves open the possibility that the partition function 
converges for the class of well-behaved pseudomanifolds considered
by Perez --- namely, those where each triangle lies in at least three 
4-simplices.   Proving this would require good bounds on the $10j$ symbol.  

Numerical computations suggest that the following bound holds:
\[  \left| \TenJ \right| \le C_1 
\prod_{i = 1}^{10} (2j_i + 1)^{-\frac{1}{5}}.\]
This is consistent with our previous observation that the $10j$
symbols decay as $O(j^{-2})$ as all spins are rescaled by the same
factor, but it gives more information when some spins are much larger
than others.  We can use this bound to sketch a rough argument 
that the partition function converges for well-behaved pseudomanifolds
in which each triangle lies in at least three 4-simplices. 

Far from the border of admissibility, it is easy to prove that
the $4j$ symbols satisfy 
\[   \left| \fourtheta{j_1}{j_2}{j_3}{j_4} \right| \ge C_2
\prod_{i = 1}^4  (2j_i + 1)^{\frac{1}{4}}   .\]
If we could ignore the border of admissibility, this estimate
and the above bound on the $10j$ symbols would imply 
a bound on the cutoff partition function:
\[
\begin{array}{ccl}
Z_J(M) &\le&  \displaystyle{C_3 \sum_{|F| \le J} 
                   \prod_{f \in \Delta_2}  (2j(f) + 1)^{-\frac{9}{20}n(f)}} \\
&& \\
       &\le&   \displaystyle{ C_3 \prod_{f \in \Delta_2} 
                     \sum_{j(f) \in \{0,\frac{1}{2}, \dots , J \}}  
                      (2j(f) + 1)^{-\frac{9}{20}n(f)}} . 
\end{array}
\]
Here $n(f)$ is the number of vertices (or equivalently, edges) of the 
face $f$, which is the same as the number of 4-simplices containing
the triangle dual to $f$.  The curious number $\frac{9}{20}$ comes
from the fact that each vertex of the face $f$ gives a factor of 
$(2j(f)+1)^{-\frac{1}{5}}$, while each edge gives a factor of 
$(2j(f)+1)^{-\frac{1}{4}}$, and $\frac{1}{5} + \frac{1}{4} = \frac{9}{20}$.
Since
\[      \sum_{j \in \{0,\frac{1}{2},\dots\}} (2j + 1)^{-\frac{9}{20}n} \]
converges when $n \ge 3$, this bound would imply convergence of $Z_J(M)$ as
$J \to \infty$ whenever each triangle is contained in at least
three 4-simplices.   Unfortunately, this argument neglects
the border of admissibility, where the $4j$ symbols grow more slowly.
Luckily, the $10j$ symbols decay more rapidly near the border of
admissibility!  We are therefore optimistic that this hole in the
argument can be fixed.

As with other versions of the Barrett--Crane model, we can get a
qualitative feel for the new model using the Metropolis algorithm.
Table 6 shows a small portion of a typical run of this algorithm, again
using the triangulation of $S^4$ as the boundary of a 5-simplex and
imposing a spin cutoff of $\frac{5}{2}$.  This table is organized just like
Tables 2 and 4.  Note that in the new model, both low spins and spins near
the cutoff show up frequently, but with a predominance of low spins.

\vskip 1em 

{\vbox{   
\begin{center}   
{\small
\setlength{\extrarowheight}{1.5pt}
\setlength{\tabcolsep}{2.2pt}
\begin{tabular}{|>{$}c<{$}|>{$}c<{$}|D{X}{\,\cdot\,}{5}|}    \hline
\text{iteration} & F          & \multicolumn{1}{c|}{$Z(F)$}\\ \hline
4995398 & 03103000002104313300 & 4.768 X 10^{-7}   \\   \hline
4995458 & 03104000001104314400 & 1.953 X 10^{-7}   \\ \hline
4995513 & 04104000000103414400 & 1.600 X 10^{-7}   \\ \hline
4995517 & 04104001000102413401 & 6.250 X 10^{-8}    \\ \hline
4995520 & 04114001000112303401 & 2.441 X 10^{-8}   \\ \hline
4995529 & 04104001000102413401 & 6.250 X 10^{-8}   \\ \hline
4995534 & 04104001100102313300 & 1.526 X 10^{-7}   \\ \hline
4995542 & 04104001200102213201 & 6.028 X 10^{-8}   \\ \hline
4995547 & 04104002200101212202 & 1.191 X 10^{-8}   \\ \hline
4995554 & 14105112200101212202 & 4.961 X 10^{-10}  \\  \hline
4995565 & 05215112200101212202 & 2.297 X 10^{-10}  \\ \hline
4995576 & 05215113200100211201 & 9.303 X 10^{-9}   \\ \hline
4995577 & 05215113100100311302 & 2.943 X 10^{-9}    \\ \hline
4995582 & 05215013100200312312 & 3.489 X 10^{-10}   \\ \hline
4995587 & 05215013110200322301 & 4.361 X 10^{-10}   \\ \hline
4995596 & 04215013111201222301 & 2.224 X 10^{-10}   \\ \hline
4995601 & 04215013211201122202 & 8.954 X 10^{-11}   \\ \hline
4995610 & 04215013311201022101 & 1.608 X 10^{-9}    \\ \hline
4995620 & 04115013311102012101 & 6.441 X 10^{-9}    \\ \hline
4995626 & 04114013310102011001 & 1.031 X 10^{-7}    \\  \hline
\end{tabular}} 
\end{center}   
\vskip 1em 
\centerline{Table 6: sample Metropolis labellings --- the new model with
spin cutoff $\frac{5}{2}$}
}}   
\vskip 1em   

It is interesting to see the frequencies with which faces are labelled
by various spins.  We show this in Table 7, based on a Metropolis run
with spin cutoff $J = 50$ and half a billion iterations. The results
obtained are very similar to results obtained with a spin cutoff of $J =
\frac{25}{2}$, indicating that they are not just an artifact of the cutoff.  
We only show results up to $j = 5$, but higher spins were seen as well,
with smoothly declining frequencies.  The most important thing to note
is that while spin zero is the most likely spin to occur, there is still
a substantial fraction of faces labelled by other spins.

\vskip 1em

{\vbox{   
\begin{center}   
{\small
\setlength{\extrarowheight}{1.5pt}
\setlength{\tabcolsep}{2.2pt}
\begin{tabular}{|>{$}c<{$}|>{$}c<{$}|}    \hline
\text{spin} & \text{frequency}  \\ \hline
 0 &   69.548 \%                  \\ \hline
 1/2 & 18.733 \%                  \\ \hline
 1 &   6.2878 \%                  \\ \hline
 3/2 & 2.5510 \%                 \\ \hline
 2 &   1.1958 \%                 \\ \hline
 5/2  &.61995 \%                 \\ \hline
 3  &  .34893 \%                \\ \hline
 7/2 & .21243 \%                \\ \hline
 4  &  .13535 \%                \\ \hline
 9/2 & .08989 \%                \\ \hline
 5 &   .06252 \%                \\ \hline
\end{tabular}} 
\end{center}   
\vskip 1em 
\centerline{Table 7: spin frequencies in $S^4$ --- the new model}
}}   
\vskip 1em   

\section{Implications} \label{implications}

Our computation of partition functions is only relevant to the `physics'
of the models being studied to the extent that it sheds light on the
behavior of observables.  The issue of observables in quantum gravity is
a thorny one, but we really need to confront it here.  One approach
would be to follow the ideas of canonical quantum gravity and use a sum
over open spin foams to compute the projection onto the space of
physical states \cite{Arnsdorf,RR,Rovelli}.  Observables would then be
described as operators on this Hilbert space.  While conceptually
well-motivated, this approach will take a great deal of work to
implement.  One of the goals of the spin foam program is to develop a
`sum over histories' approach to quantum gravity that has a chance of
making more rapid progress \cite{PerezRovelli2}.  Ideally this approach
would be compatible with the canonical approach, but not require an
explicit computation of the projection onto physical states.  In this
section we attempt to interpret our computations using this `sum over
histories' approach.

To begin with, let us tentatively call any function $O$ from spin foams
to real numbers an `observable'.   Fixing a spin foam model, we 
can try to compute the expectation value of $O$ as follows: 
\begin{equation}
    \langle O \rangle = \frac{ \sum_F O(F) Z(F)}{\sum_F Z(F) }  ,
\label{expectation}
\end{equation}
where $Z(F)$ is the amplitude our model assigns to the spin foam $F$,
and the sum is taken over all spin foams.  The denominator of
this fraction is the partition function.  

Formulas like equation (\ref{expectation}) are familiar in quantum field
theory on a fixed background spacetime, but we must reevaluate their
meaning in the current context.  Exactly what sort of `expectation
value' is this quantity $\langle O \rangle$?  In quantum field theory a
formula of this sort is used to compute vacuum expectation values.
However, in the context of quantum gravity, the notion of energy and
thus the whole concept of `vacuum' becomes problematic.  We thus propose
to interpret $\langle O \rangle$ as the average of $O$ over all
histories.  This is consistent with the interpretation of a spin foam
as a `quantum history' \cite{Baez2}.

Now, for many functions $O$ we do not expect $\langle O \rangle$ to be
well-defined.  For example, if $O(F)$ is some measure of the total
4-volume of the spacetime corresponding to $F$, there is no obvious
reason why $\langle O \rangle$ should converge.  This is not bad; it
just means that the question ``what is the expected 4-volume of
spacetime?''\ is ill-posed when one is given no further information about
the history in question.  There is no reason a theory should be able to
answer this sort of question.

However, one can often make ill-posed expectation values well-posed by
`conditioning' them.  In other words, instead of asking ``what is the
expected value of $O$?''\ one asks ``what is the expected value of $O$,
{\it given that}...?''  In physics this conditioning is usually done by
specifying either a state  or the value or expectation value of some
observables. For example, in the path integral approach to quantum
mechanics, one can compute the expected value of the position of
a particle at $t = 1$, given its position at $t = 0$, by restricting 
the path integral to paths that start at this given position at $t = 0$.   
We can even condition further by specifying information about its position 
at some other times; this has been extensively studied in the 
theory of consistent histories \cite{GH,Griffiths,Omnes}.

A path represents a classical history, but we can also do conditioning
when we compute expectation values by summing over quantum histories.
The most familiar example of a quantum history is a Feynman diagram.
Using Feynman diagrams we can compute the expectation value of some
observable measured in the future, given information about the incoming
particles in the past, by restricting the sum over Feynman diagrams to
those with certain specified incoming edges.  We can even condition on
properties of the internal edges of a Feynman diagram, e.g.\ computing
the probability that two electrons scatter given that they have
exchanged a specific number of virtual photons.  One must be careful
when working with probabilities of this sort, since they can fail to
satisfy the classical rules, thanks to interference effects.  However,
the theory of consistent histories provides a framework for correctly
dealing with them.

Spin foams are close analogs of Feynman diagrams, and indeed they {\it
are} Feynman diagrams in the group field theory approach \cite{RR2}.
This means that, as with Feynman diagram theories, in spin foam models
we can condition any expectation value by limiting the class of spin
foams to be summed over, or weighting them with a suitable factor.  This
amounts to replacing the formula for $Z(F)$ by a modified formula which
takes this conditioning into account.

The simplest example consists of setting $Z(F)$ to zero when $F$ fails
to lie in the dual 2-skeleton of a fixed triangulated 4-manifold.  While
 doing this simplifies many calculations, and we have implicitly done so
throughout this paper, it would only be physically well-motivated if we
knew spacetime were equipped with a specific triangulation.  A more
realistic example would arise if we were trying to use a Lorentzian spin
foam model to make predictions about the collision of gravitational
waves.  Here we would need to restrict the integral over spin foams to
those having a spacelike slice in which incoming gravitational waves of
a specified sort were present.  If we were studying these waves in a
bounded region of spacetime, we might also restrict to open spin foams
of a certain `size'.  Of course, we are far from being able to do this
at present.

Having modified $Z(F)$ to take the conditioning into account, there are
still some further subtleties about the convergence of formula
(\ref{expectation}).  If the sums in both numerator and denominator
converge, $\langle O \rangle$ is well-defined.  However, we can make do
with less: if both sums diverge, we can impose a cutoff on both, take
the ratio of the two, and then try to take a limit as the cutoff is
removed.  It follows that the convergence of the conditioned partition
function is neither necessary nor sufficient for computing $\langle O
\rangle$.  

What does it mean if we need to impose a cutoff and then take a limit as
the cutoff is removed to compute expectation values of observables?  The
answer depends on the nature of the cutoff, so it is good to focus on a
concrete example.  We have seen one example in Section \ref{BC}, where
we considered the DFKR model on a fixed triangulation of the 4-sphere. 
Since the partition function diverged, we found it useful to insert a
cutoff on the spins labelling spin foam faces, or equivalently,
triangles in the triangulation.  In this model we can define the 
area of a triangle labelled by the spin $j$ to be
\begin{equation}
   {\rm area} =  8\pi \gamma \ell_P^2 \sqrt{j(j+1)} ,
\label{area}
\end{equation}
where $\ell_P$ is the Planck length and $\gamma$ is an {\it ad hoc}
constant roughly analogous to the Barbero--Immirzi parameter
\cite{CMPR,Livine}.  A spin cutoff then amounts to a kind of `infrared
cutoff', since it rules out spacetime geometries containing triangles
of large area.

Without any further conditioning, the expectation value of an observable
in this theory would be given by
\begin{equation}
    \langle O \rangle = \lim_{J \to \infty}
\frac{ \sum_{|F| \le J} O(F) Z(F)}{\sum_{|F| \le J} Z(F) }  .
\label{expectation2}
\end{equation}
What does it mean when this limit exists but both numerator and
denominator diverge as $J \to \infty$?  It means that the dominant
contribution to the expectation value of the observable $O$ comes from
spacetime geometries containing triangles of arbitrarily large area!
This seems physically unrealistic, since in our world spacetime
discreteness exists, if at all, only on short length scales.  

One possible objection is that this does not take into account the
conditioning needed to phrase a sensible physical question.  Perhaps
conditioning automatically damps the contribution of spin foams with
large triangles.  This seems most plausible if we are asking questions
about a bounded region of spacetime and restrict the sum over spin foams
to those of a certain `size'.

Another way out might be to take a limit where we let the Barbero--Immirzi
parameter go to zero as we let $J \to \infty$: in other words, a kind of
`continuum limit', where spacetime discreteness gets pushed to ever
smaller distance scales.   A limit of this sort has already been
discussed by Bojowald \cite{Bojowald} in the context of Lorentzian
quantum gravity, working with the real Ashtekar variables.  He shows
that in this limit, loop quantum cosmology reduces to ordinary quantum
cosmology.  One can imagine taking a similar limit in Riemannian quantum
gravity.  However, a great deal more work would be required to see if
this is a viable strategy.

In short, the divergence of the partition function as we remove the spin
cutoff is not necessarily a disaster for the DFKR model.  However, it
seems one would need some sophisticated maneuvers to extract interesting
results from this model.  Anyone interested in this might be wise
to start by reexamining the Ponzano--Regge model of 3-dimensional
Riemannian quantum gravity, which exhibits a similar divergence.

We can avoid or at least postpone facing these subtleties by working
with the Perez--Rovelli model, where the partition function converges
for a fixed triangulation.  Of course, a divergence may still arise in
the sum over triangulations.  But still more pressing, in our opinion,
is the task of understanding spin-zero dominance.  While this phenomenon
has no obvious analogue in the Lorentzian Barrett--Crane model, it is
still worth pondering.  What does it mean when the partition function is
dominated by spin foams whose faces are mostly labelled by spin zero?
If we believe equation (\ref{area}), these correspond to triangles of
area zero!  Perhaps this model is a theory of highly degenerate quantum
geometries where most of the 4-simplices are shrunk to nothing, vaguely
reminiscent of the `crumpled phase' in Euclidean quantum
gravity~\cite{ADJ,Loll}.  Perhaps suitable conditioning will remove this
effect.  Perhaps spin-zero faces should be ignored for some reason.  Or 
perhaps Alekseev, Polychronakos, and Smedb\"ack \cite{APB} are right,
and the correct area formula is given not by equation (\ref{area}) but
by
\[
   {\rm area} =  8\pi \gamma \ell_P^2 (j + \frac{1}{2})  .
\]
This would drastically affect our interpretation of spin-zero faces.

At present, all we can say for sure is that a theory with drastic
spin-zero dominance raises as many difficult issues for our approach to
computing expectation values as one where the partition function diverges 
as the spin cutoff is removed.  The new model described in this paper seems
to avoid these issues; here we can compute expectation values of
observables and get some interesting results, at least if we fix a
triangulation.  

The simplest observable in the Riemannian Barrett--Crane model is the
average area of a triangle.  Ignoring a factor of $8 \pi \gamma
\ell_P^2$, this is given by
\[  O(F) = 
\frac{1}{|\Delta_2|} \sum_{f \in \Delta_2} \sqrt{j(f)(j(f) + 1)}  \]
where $|\Delta_2|$ is the number of triangles in the triangulation.
We use $\langle O \rangle_J$ to stand for the expectation value of
this observable with a spin cutoff of $J$:
\[  \langle O \rangle_J = 
\frac{ \sum_{|F| \le J} O(F) Z(F)}{\sum_{|F| \le J} Z(F) }  \]
where for a simple calculation we sum over spin foams in the dual
2-skeleton of the triangulation of $S^4$ as the boundary of a
5-simplex.  Some results for this quantity are shown in Table 7.  The
results for $J \le \frac{5}{2}$ are exact, while the results for higher
cutoffs are approximate, obtained using the Metropolis algorithm.  
The last data point was run using no cutoff;  spins larger than 54
never occurred in our runs, even after more than 11 billion iterations.
It appears that the limit 
\[      \langle O \rangle = \lim_{J \to \infty} \langle O\rangle_J \]
exists.  

\vskip 1em

{\vbox{   
\begin{center}   
{\small
\setlength{\extrarowheight}{1.5pt}
\setlength{\tabcolsep}{2.2pt}
\begin{tabular}{|c|c|}    \hline
$J$ & $\langle O \rangle_J$                                      \\ \hline
0     &   0.000000   \\ \hline
$1/2$ &   0.121987   \\ \hline
1     &   0.210441   \\ \hline
$3/2$ &   0.265911   \\ \hline
2     &   0.302153   \\ \hline 
$5/2$ &   0.326524    \\ \hline  \hline
$15/2$ &      0.381160   \\ \hline 
$25/2$ &      0.396701   \\ \hline 
$50$   &      0.399991   \\ \hline 
$\infty$ &    0.400005   \\ \hline
\end{tabular}} 
\end{center}   
\vskip 1em 
\centerline{Table 7: Expected average area of a triangle in $S^4$ --- 
            the new model with spin cutoff $J$}
}}   
\vskip 1em   

Now, an interesting thing about the average triangle area is that 
for this observable, the limit 
\[       \lim_{J \to \infty} \langle O\rangle_J \]
can only exist if the partition function converges.  This means our
numerical evidence that $\langle O \rangle_J$ converges is also
numerical evidence that the partition function converges!  To see why
this is true, suppose the partition function diverges.  Since in this
model the spin foam amplitudes $Z(F)$ are always nonnegative
\cite{BaezC}, the cutoff partition functions $Z_J(M)$ must approach
$+\infty$ as we remove the spin cutoff.  This implies that for any spin
$J$ we have
\[   2{\sum_{|F| \le J} Z(F) } \le  {\sum_{|F| \le J'} Z(F) } \]
for all sufficiently large spins $J'$.  This implies
\[   {\sum_{|F| \le J'} Z(F) } \le  2{\sum_{J < |F| \le J'} Z(F) }. \]
Using this, we see that for all sufficiently large $J'$,
\[
\begin{array}{ccl}
\langle O \rangle_{J'} &=& 
\displaystyle
{\frac{ \sum_{|F| \le J'} O(F) Z(F)}{\sum_{|F| \le J'} Z(F) }} \\
&&  \\          
&\ge& 
\displaystyle
{\frac{ \sum_{J < |F| \le J'} O(F) Z(F)}{\sum_{|F| \le J'} Z(F)}} \\
&&  \\          
&\ge& 
\displaystyle
{\frac{ \sum_{J < |F| \le J'} O(F) Z(F)}{2\sum_{J < |F| \le J'} Z(F) }} \\
&&   \\
&\ge& \displaystyle{\frac{\sqrt{J(J + 1)}}{2|\Delta_2|}}   \\
\end{array}
\]
since in any spin foam with $J < |F|$, there is at least one  triangle
with area at least $\sqrt{J(J + 1)}$,  so the average triangle area is
at least this quantity divided by the number of triangles.  Since $J$
can be chosen as large as we like,  we see that 
\[   \lim_{J' \to \infty} \langle O \rangle_{J'}  = +\infty .\]
This is a nice example of how the convergence of the partition function
is intimately linked to the behavior of observables.

In conclusion, we wish to emphasize that while the interpretation of
spin foam models raises many difficult issues, this is only to be
expected given the novelty of the whole setup.  We expect rapid progress,
especially if the more traditional tools of theoretical physics are
supplemented by computer calculations.  Hard numbers have a marvelous
way of making problems more concrete.

\section*{Acknowledgements} 
We thank Greg Egan and Alejandro Perez for extremely valuable
discussions.  We also thank SHARCNet for providing the supercomputer at
the University of Western Ontario on which we computed $Z_{5/2}(M)$ in
our new model.  The second author was supported by a grant from NSERC,
and the third and fourth authors were supported by NSERC and SHARCNet.

\end{document}